# Digital Divide: Mapping the geodemographics of internet accessibility across Great Britain

**Claire Powell & Luke Burns**

University of Leeds, Leeds, United Kingdom

## Abstract

**Aim:** This research proposes the first solely sociodemographic measure of digital accessibility for Great Britain. Digital inaccessibility affects circa 10 million people who are unable to access or make full use of the internet, particularly impacting the disadvantaged in society. **Method:** A geodemographic classification is developed, analysing literature-guided sociodemographic variables at the district level. **Analysis:** Resultant clusters are analysed against their sociodemographic variables and spatial extent. Findings suggest three *at-risk* clusters exist, 'Metropolitan Minority Struggle', 'Indian Metropolitan Living' and 'Pakistani-Bangladeshi Inequality'. These are validated through nationwide Ofcom telecommunications performance data and specific case studies using Office for National Statistics internet usage data. **Conclusion:** Using solely contemporary and open-source sociodemographic variables, this paper enhances previous digital accessibility research. The identification of digitally inaccessible areas allows focussed local and national government resource and policy targeting, particularly important as a key data source and methodology post-2021, following the expected final nationwide census.

# Introduction

Digital accessibility refers to the non-egalitarian divide between those with and those without internet access (Castells, 2002; Singleton, et al., 2020). Overall, inaccessibility levels have declined steadily in recent years, but a considerable proportion of the Great Britain population remain without access (circa 10 million people in 2019 (Blank, et al., 2020) (Dutton & Blank, 2013)). Even with internet access, a minimum speed and connectivity is required to enable multiple users using the same internet connection to carry out common daily tasks (UK Government, 2020a), many of which are taken for granted.

According to the Office of Communications (Ofcom) and the UK Government (2020a) under the Broadband Universal Service Obligation, a 'decent broadband service' has an upload speed of 1 Mbit/s (megabit per second) and download speed of 10 Mbit/s. When compared to Europe as a whole, the UK ranks 5$^{th}$ in internet services use (European Commission, 2020).

Digital accessibility can be split into three aspects: The first-level digital divide is the ability (or inability) to access the internet due to physical infrastructure or financial constraints. The second-level digital divide is how effectively people engage computer skills to exploit internet benefits (Hargittai, 2002; van Deursen & van Dijk, 2011). The third-level digital divide addresses internet use consequences (Selwyn, 2004; van Deursen, et al., 2017). This research crosscuts all aspects, with a key focus on the first and second levels.

Factors influencing nationwide digital accessibility are multidimensional, often interrelated and are generally directly associated with sociodemographic attributes that vary spatially. Geodemographics, regularly defined as "the analysis of people according to areas where they live" (Sleight, 1997, p. 16), takes into account socio-economic and demographic similarities and differences, and has had considerable public and private sector success (Harris, et al., 2005; Webber & Burrows, 2018).

In this research, the first solely sociodemographic measure of digital accessibility for Great Britain will be presented, with findings analysed and validated. The most recent, freely available administrative data and open-source software will be used, ensuring that the output is both transparent and easy to replicate and update. Results are likely to aid future policy recommendations pertaining to internet accessibility (and inclusivity) and influence more broad internet connectivity debates.

# Research Rationale

Geodemographic measures surrounding the topic of digital accessibility are limited but include a nationwide geodemographic 'Internet User Classification' created by Singleton et al (2020). This work made use of exclusively transactional data, derived between 2013 and 2016. In other work, Blank et al (2017) used small area estimation from individual-level small scale survey data to determine the influence of demographic characteristics versus spatial differences on 2013 internet use.

This research builds on the work of Blank et al (2017), by creating a new bespoke classification, working with more recent and freely available data and using more complete, nationwide survey data collected and verified by the UK Government. Longley et al (2008) used well-known commercial geodemographic classification MOSAIC (Experian) to determine individual levels of engagement with electronic technologies and products for marketing. This research focusses specifically on sociodemographic factors as opposed to the narrow economic influence. Sociodemographic factors encompass multiple dimensions of the population (including direct and indirect impacts). Thus, this research will further develop knowledge and contemporise past works in the domain.

Currently, no existing work has explored digital accessibility using only sociodemographic variables across Great Britain. A core focus is on the less well researched first and second digital divides. This research sets out to identify those at greatest risk of not gaining maximum benefit from the internet.

Digital accessibility research is increasingly important globally, with the plethora of data generation and technological advances. The 5G network is a wireless mobile network with increased bandwidth (Médard, 2020), higher data capacity and faster, more reliable (99.9%) latency (time between data transfer) than its predecessor 4G (Ilderem, 2020). Since May 2019, 5G has been progressively set up across Great Britain. Banning new Huawei technology will slow and redact some 5G access in the short term, however, 5G is still set to hold the near future of internet connectivity (National Cyber Security Centre, 2020; UK Government, 2020b). This nationwide digital accessibility research will help identify the types of people at greatest impact of loss of digital access and their spatial extent, particularly important as a data source and methodology to follow after the last UK census in 2021. Use of government administrative data in this research could highlight accurate district-level data sources for data analysis at national and local level for future policy decisions and target setting.

# Review of Digital Accessibility

A thorough review of past related research was undertaken, exploring the current understanding and knowledge linked to sociodemographic factors. Sociodemography is the study of groups of the population that share characteristics (Lenormand, et al., 2015). Here, variables that demonstrate fundamental disadvantage, both socially and demographically, can help highlight those potentially most *at risk* from digital inaccessibility and exclusion.

Vicente and Gil-de-Bernabé (2010), alongside Epstein et al (2011), and White and Selwyn (2013), suggest a more nuanced understanding of digital accessibility issues (backed by a collective societal responsibility) could aid future government digital policies. This could also aid advocacy of the internet as a utility, a service the population cannot live without thus short-term faults merit immediate repair. This is popular with many (e.g. Skerratt et al (2008), Townsend et al (2013) and Philip et al (2015)) and already present in countries such as Sweden and Finland.

Selwyn et al (2005) carried out an adult-focussed internet usage interview study which found sociodemographic factors influencing internet-derived knowledge and in turn impacting employment applications, job progression, health enquiries, social communication, business operations and more. Blank et al (2020) also analysed multiple demographic factors (age, education, income, functional literacy, gender, employment status, marital status, social grade, ethnicity, children in household, disability) to determine nationwide internet use in the 2019 Oxford Internet Survey.

## Qualifications

The most common digital access barrier identified in past work was educational qualifications/attainment. Such qualifications indicate likely digital skills, ambition and opportunities. A Dutch Internet Benefits survey by van Deursen and Helsper (2017) found well-educated individuals are likely to be connected and regular internet users with advanced internet skills and a greater quantity of positive digital experiences. Dutton and Blank (2013) support this in a UK context through the Oxford Internet Survey, with 95% of university graduates online compared to 40% with no educational qualifications. Those with graduate-level qualifications, in line with National Vocational Qualification (NVQ) 4, and higher are more adept at grasping online opportunities than those less qualified (Blank & Lutz, 2016).

The type of internet usage differs by qualification. Drawing from the Bourdieu (1977) Cultural and Social Reproduction theory, Weber and Becker (2019) found well-educated European adolescents use the internet more for school and educational work than entertainment. Additionally, well-educated parents (particularly those using IT at work (Mesch & Talmud, 2011)) encourage higher level IT activities (e.g. website creation) than their less educated, less supported peers. A supportive environment promotes internet exploration, experimentation and IT skills gain (Weber & Becker, 2019).

Rural digital research by Townsend et al (2013) noted UK-wide influences on internet access, finding educated adults also gain from online employment opportunities and career development, this being fundamental in a fluctuating economy where increasing human capital through adaptability and retraining is vital to remain employable. Longley et al (2008), through the UK e-Society National Classification, noted basic IT skills (defined in van Deursen et al (2016) internet skills framework as the ability to find, select, and evaluate online information) are considered employee responsibility and IT skills are as important as having higher educational qualifications. Without internet, those who are less educated are restricted from educational, employment and career development opportunities and limited in their employment potential (typically gaining unskilled, manual work) (Townsend, et al., 2013).

## Employment Status

A further factor identified as influencing digital access is employment status, often linked to education and income. Higher qualifications often lead to higher salaried employment with more disposable income for broadband connectivity and internet-enabled devices. High-income users tend to carry out capital-enhancing activities and are more expressive internet users (Blank & Lutz, 2016). In contrast, those residing where education inequality exists tend to have income differentials and social disparities are perpetuated (Holsinger & Jacob, 2009).

Blank and Lutz's (2016) research on internet benefits and harms in Great Britain built on Blumler and Katz (1974) 'Uses and Gratifications' theory to reveal internet use satisfaction. Findings showed young, highly educated, high income users benefit most from digital access followed by the elderly. Although the benefits are subjective, the Uses and Gratifications theory categorises benefit levels (e.g. goal-oriented use, fulfils needs and self-aware of reasons for use). Education-income benefits are not definite with different socioeconomic groups having different needs. Some benefit educationally, others by income, and many through both (Blank & Lutz, 2016).

Xiang et al (2018) challenged findings suggesting education-income benefits do provide equal benefit with education, a poverty reduction catalyst, promoting social mobility and producing a skilled workforce. This paper focussed on Central Beijing education inequality. Other research focussing on the employment, education, and internet accessibility link include: Milanovic (2016), building on earlier work by Piketty and Saez (2003) who took a different approach, proposing the Kuznets waves theory, where technology developments, globalisation and public policy cause income inequality fluctuations. Milanovic (2016) provided evidence of the transfer from manufacturing to skill-heterogenous services (e.g. Big Data) causing rising inequality. Links between employment, education, and internet accessibility are established (Blank & Lutz, 2016; Milanovic, 2016). However, the majority of research evidences these variables as providing varying benefit.

## Age

Age can measure different life experiences, skills and knowledge. Longley et al (2008) noted young adults gained internet experience when susceptible to learning and able to afford technology. Livingstone and Helsper (2009) developed this further in their teenager-based internet skills and self-efficacy study. Internet introduction in one context, such as work or education, boosts spare time internet usage and exploration. Older community members and all disadvantaged groups tend to have fewer internet-enabled devices and lower broadband connectivity (Townsend, et al., 2014; Blank, et al., 2020). Most not having grown up using digital devices have less internet experience, lowering online skills (instrumental rather than experimental use) (Longley, et al., 2008). Fewer opportunities are grasped, potentially adding financial and health burdens. Web products have efficient supply chains, cutting costs (saving the UK £18 billion in 2009) (Kalapesi, et al., 2010). Online official health advice can promote late-life wellness (Hargittai, et al., 2019). Online social or work network exclusion can marginalise those unable to keep updated (Longley, et al., 2008).

## Ethnicity

Multiple global ethnicity-focused internet access studies, including Chen and Wellman (2004) and Mesch and Talmud (2011), found ethnic minorities tend to report less internet access. More minority workers are employed in manual jobs, where internet exposure and learning IT skills are deemed less important and are unsupported. Blank et al (2020) reinforced these findings, adding UK minorities are more likely in disadvantaged groups. Scheerder et al (2017) researched determinants of internet skills from 126 global journal articles and found preconceived negative judgements of minority groups resulting from their disadvantaged position, a leading factor in lack of internet confidence and a disincentive to internet usage.

# Research Approach

This paper adopts a seven-step structure to developing a geodemographic area-level classification, similar to that proposed by Gibson and See (2006) and Burns et al. (2018). Figure 1 summarises each of the seven replicable phases and the discussion that follows provides additional contextual information.

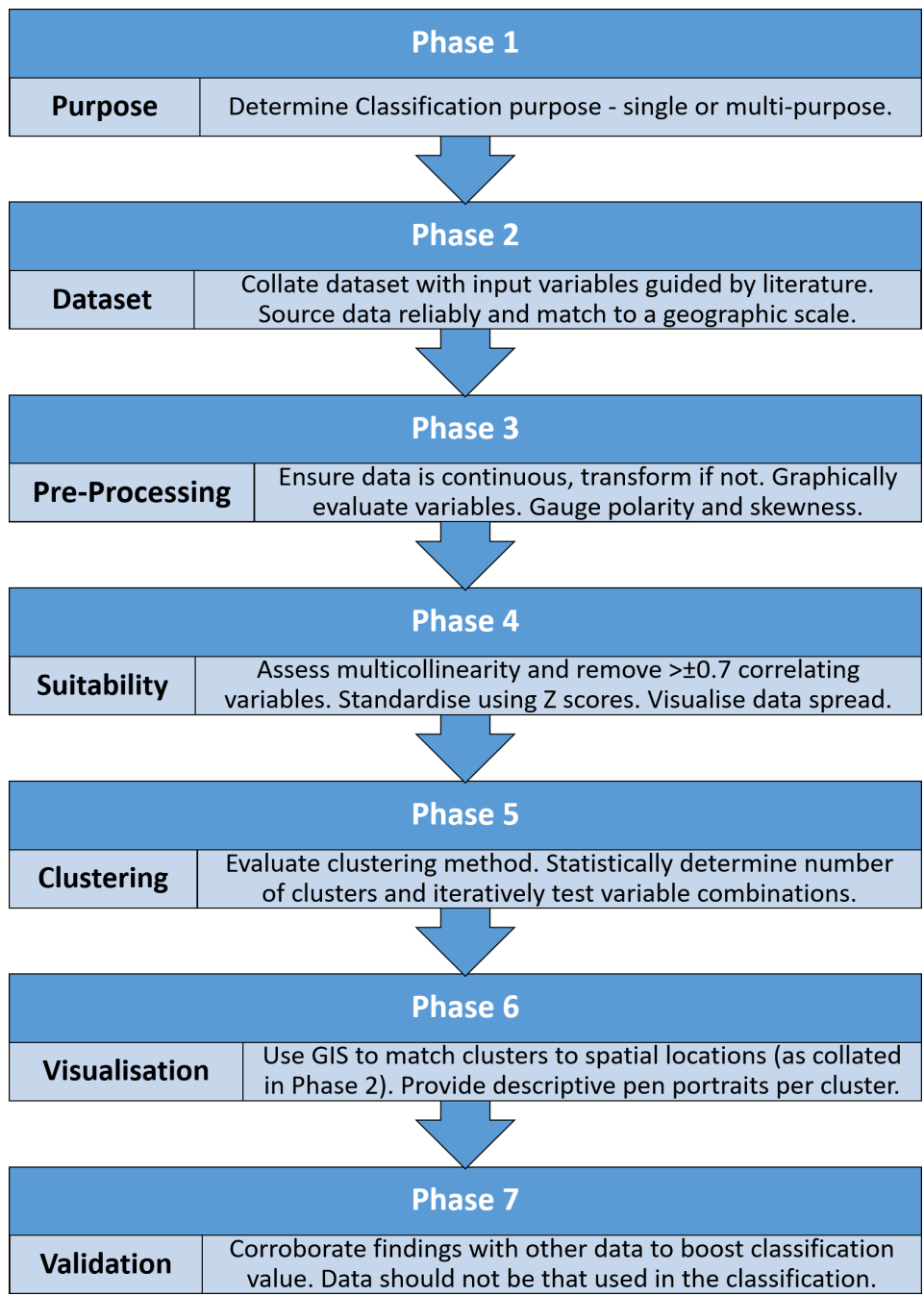

**Figure 1** Geodemographic System Framework, adapted from Gibson and See (2006, p. 214) and Burns et al (2018, p. 421).

# Phase 1: Purpose

The classification put forward in this research is the first solely sociodemographic measure of digital accessibility for Great Britain. Guided by literature intelligence, the variables presented in Table 1 were determined as being most effective at showing districts *'at risk'* from digital inaccessibility. These can be divided into two broad categories: 'Demographic' and 'Social'.

Table 1: Domains, Dimensions and Measures of Digital Accessibility. Adapted from Singleton et al (2020). Data derived from the ONS Annual Population Survey (2019a).

| Domains | Dimensions | Measures |
|---|---|---|
| Demographic | Age | 16-24 years<br>25-34 years<br>35-44 years |
|  | Ethnicity | Mixed<br>Indian<br>Pakistani/Bangladeshi<br>Black<br>Other Minority Ethnicities |
| Social | Qualifications | National Vocational Qualification (NVQ) 3+ (4/5 = e.g. Undergraduate, 6 = Graduate, 7 = Postgraduate, 8 = Doctorate) |
|  | Employment Status | Unemployment<br>Inactive |

# Phase 2: Dataset

All data used in this research are open source, administrative, cover Great Britain (in full) and are derived from the Office for National Statistics (ONS) Annual Population Survey (APS) (2019a), Ofcom Telecom Operators Performance (2019a) and the ONS (2019b) Opinions and Lifestyle Survey. APS is a household survey collected every quarter with circa 320,000 respondents (0.5% of the UK population). Data are added from the Labour Force Survey and when combined have the largest coverage of any UK household survey, allowing intergenerational statistics for small geographic areas. The survey estimates key social and labour market variables (including employment and education) at local authority level (ONS, 2012). Data for January to December 2019 were used as part of this research (ONS, 2019a).

Ofcom is the UK regulator for communications. Performance data comprises broadband and mobile 4G, 3G and 2G networks, from four mobile network operators (EE, O2, Three and Vodafone) with the largest UK coverage (Ofcom, 2019b). Annual data releases are published alongside a report analysing current state of the UK communications infrastructure (Ofcom, 2019c). Ofcom uses Ordnance Survey (OS) AddressBase® Premium to assess broadband coverage of individual in-use properties, which contains 44 million addresses from Local Government National Land and Property Gazetteer, OS MasterMap address layer and Royal Mail Postal Address File (Ofcom, 2019b). District codes and raw upload, data usage and download data analysed are accurate to May 2019, with local authority boundary changes as of December 2019 (Ofcom, 2019a).

The ONS Opinions and Lifestyle Survey is part of the Labour Force Survey and has collected a wide range of UK-wide annual population data since 2011 (ONS, 2019c), with an average of 40,000 respondents. Here, data selected relate to internet usage or non-usage. Data are to the European Union NUTS (Nomenclature of Territorial units for Statistics) scale and use 2016 level 3 geographic boundaries, some of which cover the same geographical area as districts. This internet usage data is from January to March 2019 (ONS, 2019c).

Use of opensource data alongside the following clear, detailed methodology enables scientifically reproducible research to be open to scrutiny (Singleton & Longley, 2009). Non-census data allows frequent area analysis between decadal censuses and post-2021 (Leventhal, 2016). Recent and regular data releases allow funding and resource targeting of most spatially and temporally relevant results (Singleton, et al., 2016). Specifically, administrative data can generally provide a wide variety of data that can be mined to extract potentially useful information (Singleton & Spielman, 2014).

Data are aggregated to Local Authority District (317 in England), Unitary Authority (22 in Wales) and Council Areas (32 in Scotland) level (as of December 2019) (Blank, et al., 2017). Local Authority Districts vary between 2,300 and 1.1 million people, and include metropolitan districts, London boroughs, non-metropolitan districts and unitary authorities. Welsh Unitary Authority populations vary between 90,000 and 370,000 people, and Scottish Council Areas vary between 22,000 and 650,000 people (Blank, et al., 2017). Aggregation level is referred to as districts hereafter.

# Phase 3: Pre-Processing

Following variable identification from a thorough review of contemporary academic literature, 11 variables influencing digital accessibility were collated. The number of variables were limited to 11 to avoid noise, prevent misrepresentation of districts and reduce inaccuracies from poorly fitting districts in clusters (Vickers & Rees, 2007).

Prior to analysis, multiple pre-processing steps were undertaken to ensure data suitability, using opensource software, R. Where the number of survey responses was below 500, data was omitted to preserve confidentiality. This was only the case for identifying the number of survey respondents who were of minority ethnicity status. However, total ethnicity was available so missing values were able to be calculated in most cases. Where multiple data were missing, averages were calculated following known data subtraction. Data were transformed onto a continuous scale suitable for classification and polarity determined (Leventhal, 2016) (Riekkinen & Burns, 2018).

## Phase 4: Suitability

At this stage, data were assessed for suitability with regards to their inclusion in the classification algorithm. Multicollinearity was evaluated (Figure 2) with highly correlating variables further explored and consequently removed if variable impact was deemed less important than any compounding impact on results. Retaining just one variable of a highly correlating pair of variables enables each variable to contribute its own unique dimension in the geodemographic classification (Lucy & Burns, 2017). An arbitrary multicollinearity threshold of $>\pm0.7$ was selected in line with past academic research (Judge, et al., 1982; Halkos & Tsilika, 2018).

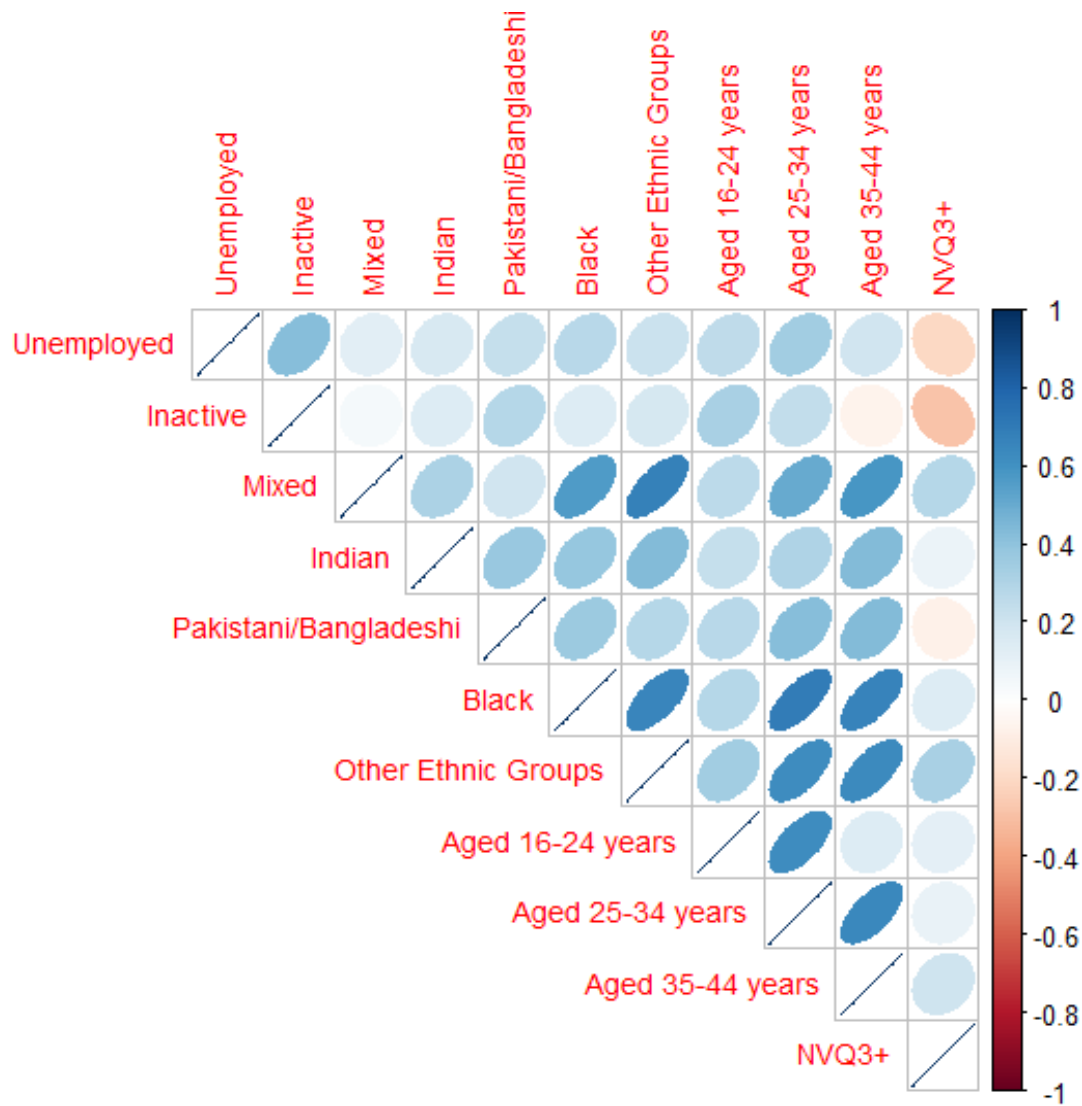

**Figure 2** Multicollinearity – Final Eleven Variables

In Figure 2, multicollinearity is observed to ±0.7 and analysed using R. Blue indicates positivity correlating variables. Red indicates negatively correlating variables. The darker the colour, the stronger the correlation. Most variables are positively correlating with only Aged 35-44 years and Inactive (employment status), NVQ 3+ and Unemployed, and Inactive and Pakistani/Bangladeshi correlating negatively. The highest positive correlation is 0.695 (Black and Aged 25-34 years); the highest negative correlation is -0.283 (Inactive and NVQ3+). All variables correlate to 0.01 with 99% statistical significance and thus can be deemed high confidence.

Variables undergo normalisation by Z-scores therefore allowing direct comparability (Vickers & Rees, 2007). Percentage values are transformed into Z-scores by: $Z = \frac{x - \mu}{\sigma}$. Z indicates the standard score, $x$ is the observed value, $\mu$ is the sample mean and $\sigma$ is the sample standard deviation (Milligan & Cooper, 1988). Z-scores do not produce normalised data with consistently the same scale. Adjustment of the scale enables an effective spread of outliers and non-outlier data to be presented (Shinwell & Cohen, 2020). For this nationwide research where there is a large dataset and outliers are likely, Z-scores tend to perform most effectively when compared to alternative approaches, such as Principle Components Analysis or Min-Max (0-1) scaling.

# Phase 5: Clustering

This research opted to use K-Means classification as the route to partition the multidimensional dataset. K-Means is an unsupervised, hard (crisp) partitioning clustering algorithm that uses machine learning to group large volumes of data based on variable similarity (MacQueen, 1967; Hartigan & Wong, 1979). K-Means has its starting seeds and the number of clusters predetermined (Major, et al., 2018). Starting seeds highly influence final cluster solution, therefore repeat-clustering (with randomised seeds) ensures more accurate and valid results (Burns, 2017; Xiang, et al., 2018). K-Means is computationally fast, accurate and sensitive to outliers (Cardot, et al., 2012; Gupta & Panda, 2018).

Guided by the literature, statistical R package, clValid, and similar previous commercial and academic success, K-Means was deemed the most appropriate algorithm for the dataset. Previous successful K-Means classifications (in the geodemographics domain) include the UK 2011 Area Classifications (ONS, 2016), Personicx by Acxiom (2020) and Experian MOSAIC system (2020). The clValid package compares multiple algorithms to identify the 'best' clustering method (Brock, et al., 2008).

Following K-Means selection, the number of clusters required identification. Statistical algorithms, including Gap Statistic and Clustergram, determine cluster numbers in R. All evaluate the whole dataset (globally) rather than analysing individual pairs of clusters (locally) to test if amalgamation improves clusters (Gordon, 1999), and all test a range of different cluster numbers and are well suited to the 9,646 data points (de Amorim & Hennig, 2015). The Gap Statistic explores partitions in the dip of a normalised performance plot. A suitable number of clusters is achieved when the smallest number of clusters (where gain is not higher than expected on the normalised performance curve) is identified (Tibshirani, et al., 2002). The Gap Statistic works well alongside K-Means. Clustergram, also effective for non-hierarchical clustering, plots a series of potential cluster frequency values alongside the weighted mean of their first principal component (Wierzchoń & Kłopotek, 2017). The resultant graph shows the relative separation of clusters. Distinctive and well-separated clusters are deemed most suitable given their homogenous nature (Schonlau, 2002).

In this research, the number of clusters was most accurately determined by using the Gap Statistic approach (Tibshirani, et al., 2002) (Figure 3) and Clustergram (Schonlau, 2002) (Figure 4). The Gap Statistic was run 500 times and Clustergram repeated 100 times, both with different cluster 'starting points' each time, ensuring initial seeds were randomly allocated and different combinations could be created (Singleton, et al., 2020).

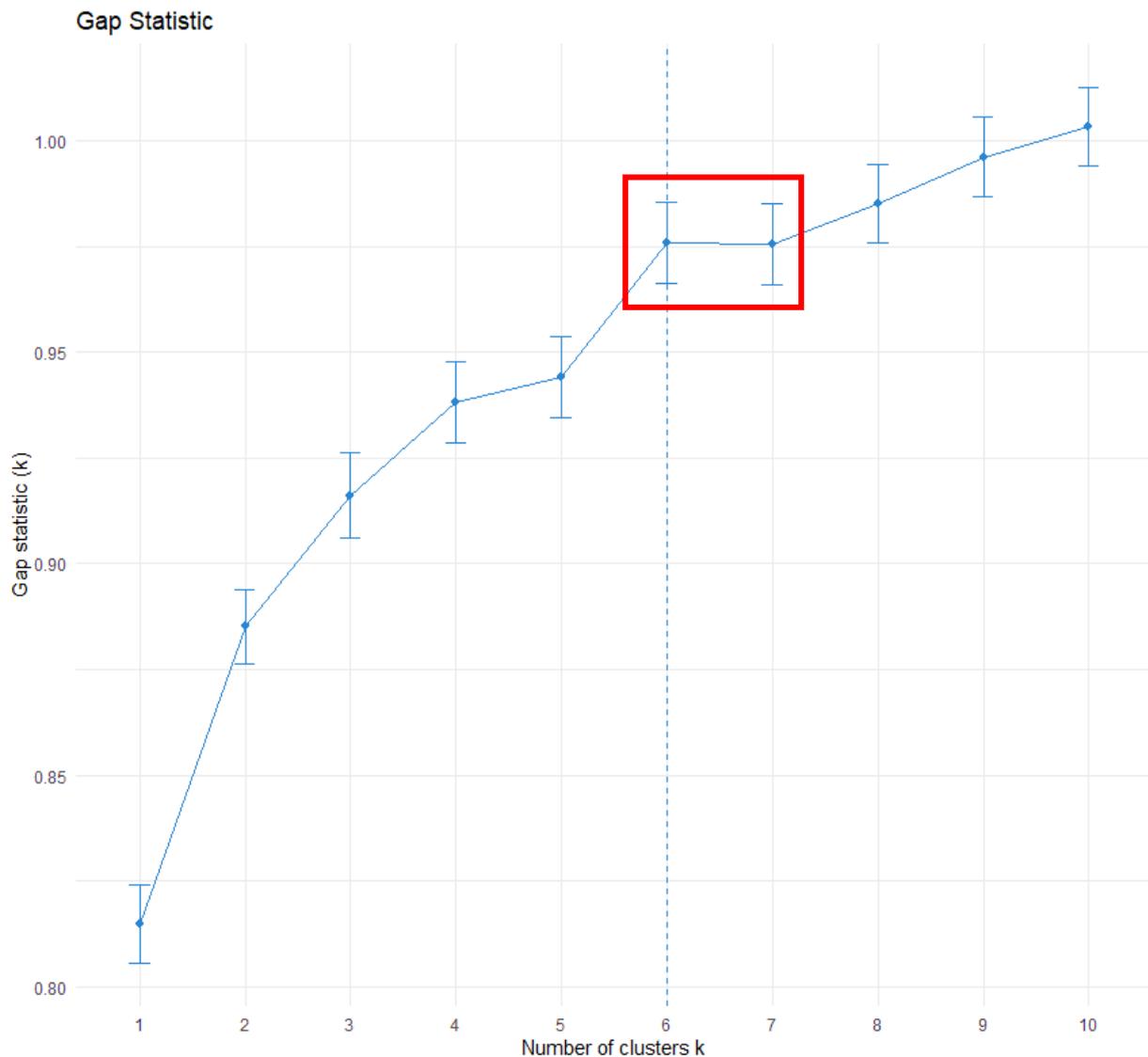

**Figure 3** Example Gap Statistic

After being run 500 times, Figure 3 suggested 6 clusters as most suitable for those specific variables. Seven clusters also appear viable with a very similar gap statistic (~0.975) and confidence intervals.

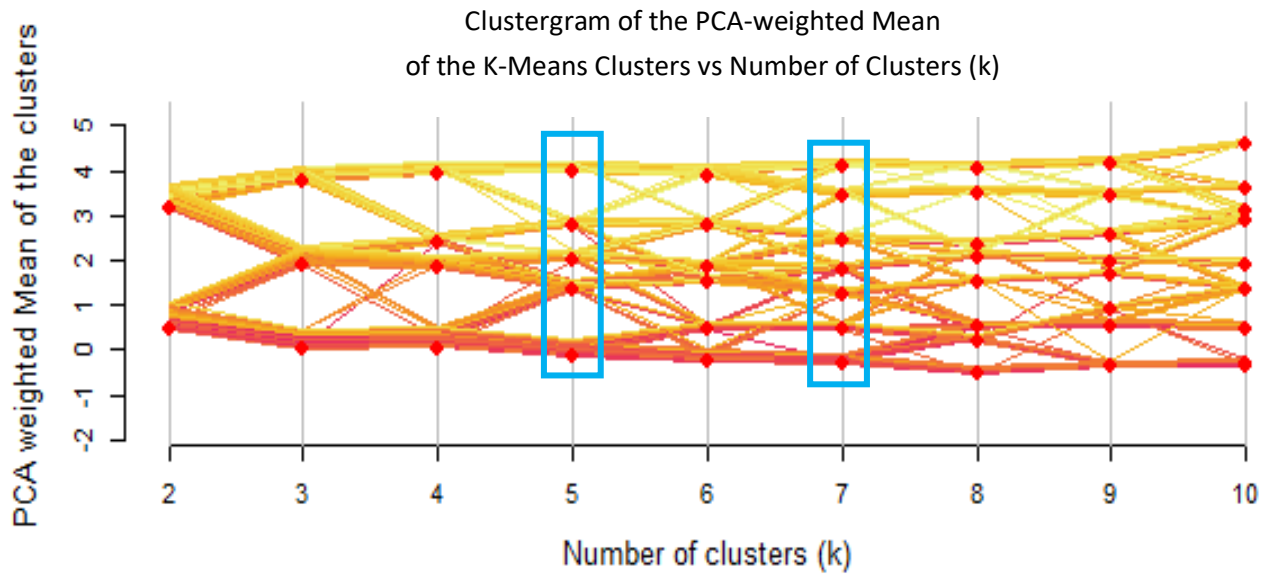

**Figure 4** Example Clustergram

From Figure 4 and all subsequent repeats, 5 and 7 clusters are consistently the two highest and most well distributed number of clusters. Seven clusters appear in both statistical methods as a potential optimum number of clusters. Although, 7 is not the optimum chosen by the Gap Statistic, values are very similar to the optimum, 6, which would not be selected as optimum by the Clustergram due to poor, inconsistent distribution of Principal Components Analysis weighted cluster means.

A seven-cluster classification was deemed most statistically suitable. Greater cluster numbers decrease cluster-cases association strength, cases are less representative of clusters (Harris, et al., 2005). K-Means involves an iterative process of moving one district (hereafter case) from one cluster to another to evaluate if a move enhances the sum of squared deviations within each cluster (Aldenderfer & Blashfield, 1984; Burns, et al., 2018). Cases are allocated (or re-allocated) to clusters until all cases are stable in clusters and provide maximum improvement to the cluster. Cases with similar variables group together and dissimilar variables exist in different clusters (Kaufman & Rousseeuw, 1990). Clusters created should represent discrete categories and reflect similar districts (Spielman & Thill, 2008). Starting cluster centres and initial seeds can determine different solutions depending on data order, therefore most accurate cluster solution requires running of the algorithm multiple times (here 1,000) with different initial cluster centres each time (Singleton, et al., 2016).

### K-Means Evaluation

K-Means attempts to minimise 'within' cluster variability and maximise 'between' cluster variability (Vickers & Rees, 2007). Cluster centres and Analysis of Variance (ANOVA) F value gauges distinctiveness and robustness of cluster-cases fit (Everitt, et al., 2011). Final cluster centres show all classification variables. Values above zero show a variable is above the population mean in the districts within that particular cluster. Values sub-zero, below the population mean, show variables that are less prevalent in those districts for that particular cluster. ANOVA F values reflect variables which provide greatest contributions to resultant clusters, as shown in Table 2, thus highlighting effectiveness of clusters. Higher F values indicate greater influence in dividing districts into separate clusters. Number of cases (or districts) in each cluster also reflects cluster effectiveness in Table 3. Evenly spread values show clusters represent a range of variable characteristics experienced in districts across Great Britain.

Boxplot showing distance of cases from cluster centres against cluster numbers shows the relative allocation of districts into clusters. The boxplot (Figure 5) is the final visualisation of SPSS cluster effectiveness. Data points residing further from the mean (or outliers) show districts that fit less suitably into clusters. In such clusters, variables are likely to be misaligned with the majority of districts in that cluster. However, a district can be placed there as it is the 'best fit' of clusters available. A classification containing many outliers across multiple clusters may suggest re-running the K-Means algorithm with more clusters may be beneficial.

## Phase 6: Visualisation

Upon K-Means completion, district codes and cluster numbers were mapped in a geographical information system (using opensource QGIS). London was mapped separately to Great Britain showing its distinct geographical patterns. Clusters were described using pen portraits – short summaries of distinctive sociodemographic and spatial features in each cluster.

## Phase 7: Validation

The classification was validated by matching case (district) codes to Ofcom (2019a) broadband performance dataset and the ONS Internet Users (2019b) dataset. Nationwide Ofcom upload and download data were compared against the Great Britain classification results to see actual areas where low upload and download speeds were present. ONS (2019b) Internet Usage data validated specific case studies representing each of the clusters. Linking the final classification to other datasets also corroborates classification success given how additional data can improve discrimination between districts (Longley, et al., 2008). Areas where download or upload speeds are slow or where internet usage is already low could identify at-risk internet inaccessibility areas.

A K-Means geodemographic classification was deemed most suitable for the 11 literature-guided sociodemographic variables. Overall statistics, the Gap Statistic and Clustergram, assessed the optimum number of clusters as 7. Greater London, known as being geographically and socio-demographically distinctive, was mapped separately to Great Britain. Methods were run hundreds of times to ensure reliable results. Findings were mapped to show spatial extent and allow validation with datasets from well-regarded organisations, Office for National Statistics and Ofcom.

# Analysis: Digital Accessibility Classification

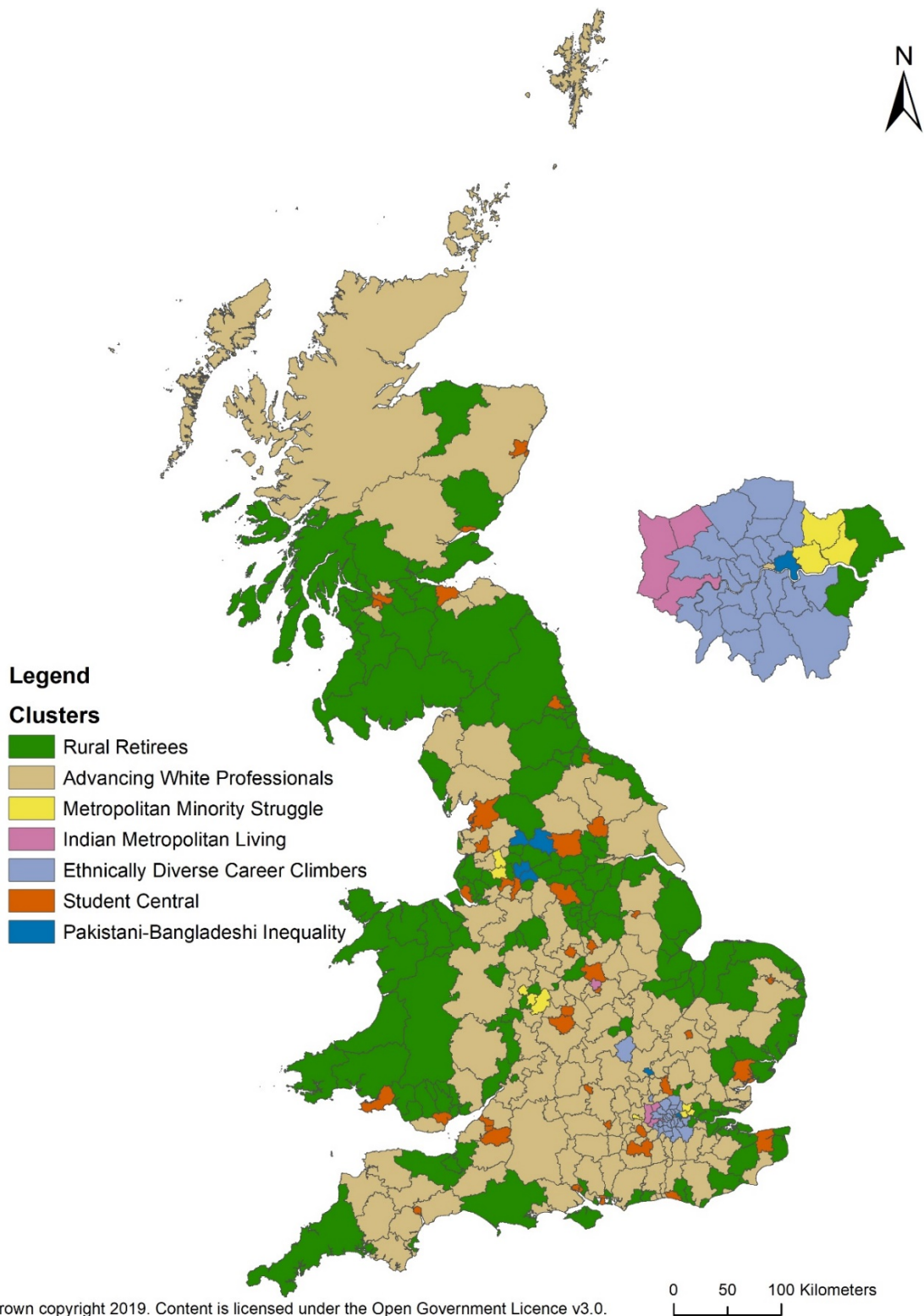

**Figure 6** Great Britain Digital Accessibility 2019 Classification Map with Greater London inset

The final digital accessibility classification (Figure 6) has nationwide and district level differences. Most 'Rural Retirees' are coast-based, whilst all other clusters tend to be inland. 'Rural Retirees' and 'Advancing White Professionals' cover the majority of Great Britain whilst 'Student Central' covers all major university cities, including Glasgow, Leeds and Warwick. Regional cluster groupings are present, particularly in Wales, the English-Scottish border and London. At district level, some are grouped with others nearby, notably the largely Northern England at-risk 'Pakistani-Bangladeshi Inequality' cluster and the 'Ethnically Diverse Career Climbers' cluster. Other clusters appear individually at specific locations such as the 'Student Central' cluster, 'Metropolitan Minority Struggle' and 'Indian Metropolitan Living'.

London is known to be socio-economically and demographically distinct to the remainder of the country, hence why London is mapped separately. Dean et al (2012) found London and the South East to be the most prosperous and 'wired' parts of the country with regards to connectivity, and later research by Dutton and Blank (2013) supported this, finding that regional Internet use varies from 60% in the North East, 71% in Wales to 86% in London and 83% in the South East, in the 2013 Oxford Internet Survey. In Figure 6, London is very clearly distinct in the classification, containing the majority of 'Indian Metropolitan Living' and 'Ethnically Diverse Career Climbers' clusters.

The Great Britain population is a mix of different sociodemographic groupings and divides (ONS, 2019a), and thus the classification provides a good district level representation. District level data also matches the level of UK Government statistics; therefore, results are easily transferable into policy (Grupp & Mogee, 2004) and match census results, potentially providing further future validation. District level is a finer spatial resolution than other coarse scales, e.g. Nomenclature of Territorial Units for Statistics (NUTS) level where large significantly different socioeconomic and demographic cities are merged (Longley, 2012). However, it is important to note that no spatial scale is ever representative of every individual. Presumption of individual characteristics about where they live from large scale results defines the ecological fallacy (Voas & Williamson, 2002). Classification outputs repositioned out of their original spatial context fall into the modifiable areal unit problem (Openshaw & Wymer, 1995). Policy leaders must be aware of aggregated data, underlying variables or spatial patterns at the original spatial extent when creating policies, to avoid potentially inaccurate presumptions (Riekkinen & Burns, 2018).

# Pen Portraits

Presumed below are pen portrait (qualitative descriptions) of each of the seven clusters as shown in Figure 6.

## Cluster 1 - Rural Retirees

All groups are based on literature-guided sociodemographic variables from Table 1. This group (Figure 7) has higher majority white ethnicity, generally based in coastal and rural areas with a large share of residents > 44 years old, inactive and unemployed. The proportion of those of Pakistani-Bangladeshi and Indian ethnicity is lowest here and in the Advancing White Professionals cluster. The majority of Great Britain aligns in Rural Retirees and Advancing White Professionals clusters. Rural retiree areas include much of Wales, the Scottish-English border, Norfolk-Suffolk coast and Cornwall.

## Cluster 2 - Advancing White Professionals

The Advancing White Professionals group (Figure 8) comprises of middle-aged working professionals typically aged between 35 and 44 years old, and above. This cluster has an above average proportion of well-educated residents with > NVQ3+ qualifications residing inland and a large proportion active or employed. The majority of Great Britain fall into this and the Rural Retirees cluster. Advancing White Professionals areas include the Scottish Highlands and Islands, and much of central Southern England.

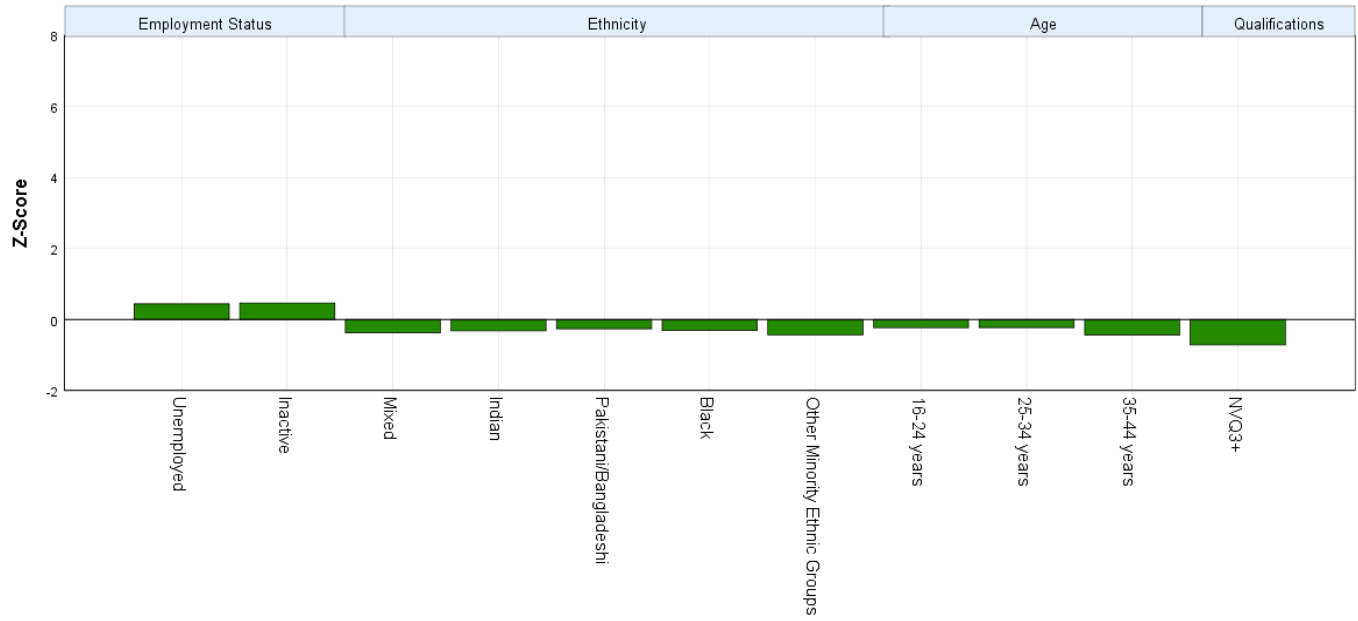

**Figure 7** Rural Retirees Profile

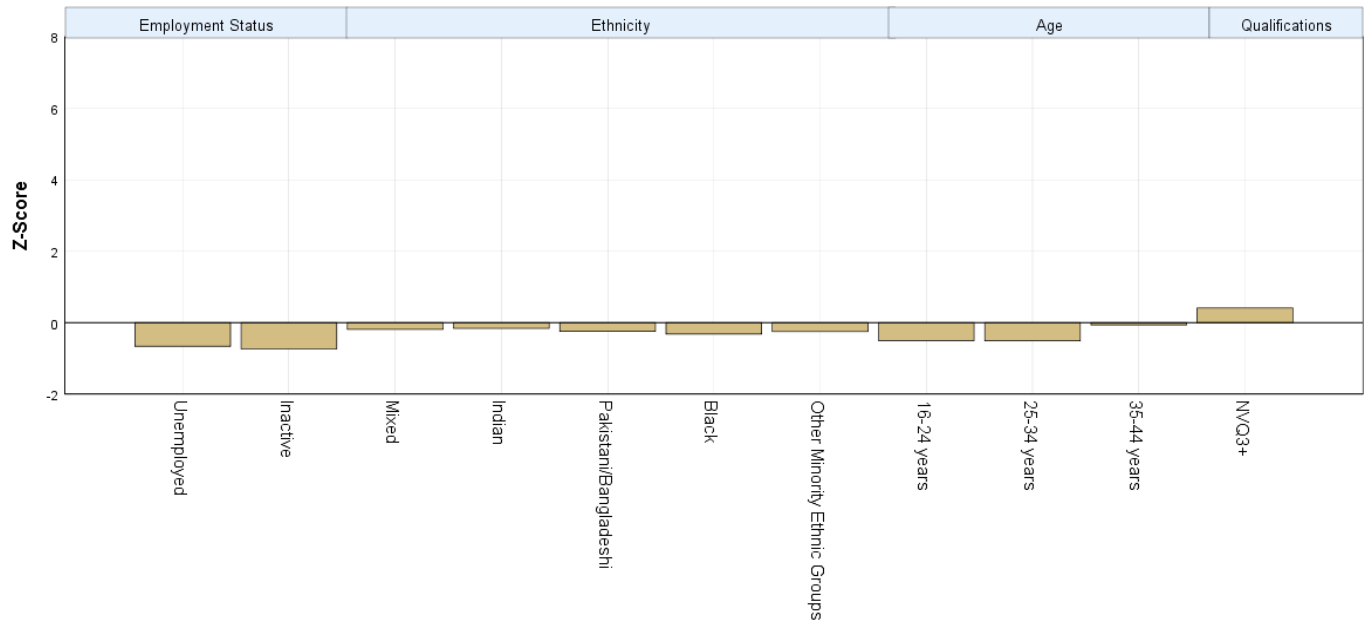

**Figure 8** Advancing White Professionals Profile

## Cluster 3 - Metropolitan Minority Struggle

This group (Figure 9), scattered in small clusters across England, are at risk of digital inaccessibility. Pakistani/Bangladeshi, Black and Indian ethnicities are above average in the population. Residents are more likely aged between 25 to 44 years old, currently inactive or unemployed with qualifications lower than NVQ3+. Metropolitan Minority Struggle areas include Birmingham, Wolverhampton and North East London.

## Cluster 4 - Indian Metropolitan Living

Within this group (Figure 10), the highest proportion of residents are of Indian ethnicity with a high proportion also from other minority ethnic groups (other than Mixed, Black and Pakistani-Bangladeshi), with similar proportions to the Ethnically Diverse Career Climbers cluster. Many in the group are aged 35-44 years old and are inactive in their employment searching. This cluster will likely also require digital access help. Indian Metropolitan Living areas include Southern England and include Hillington, Leicester and West London.

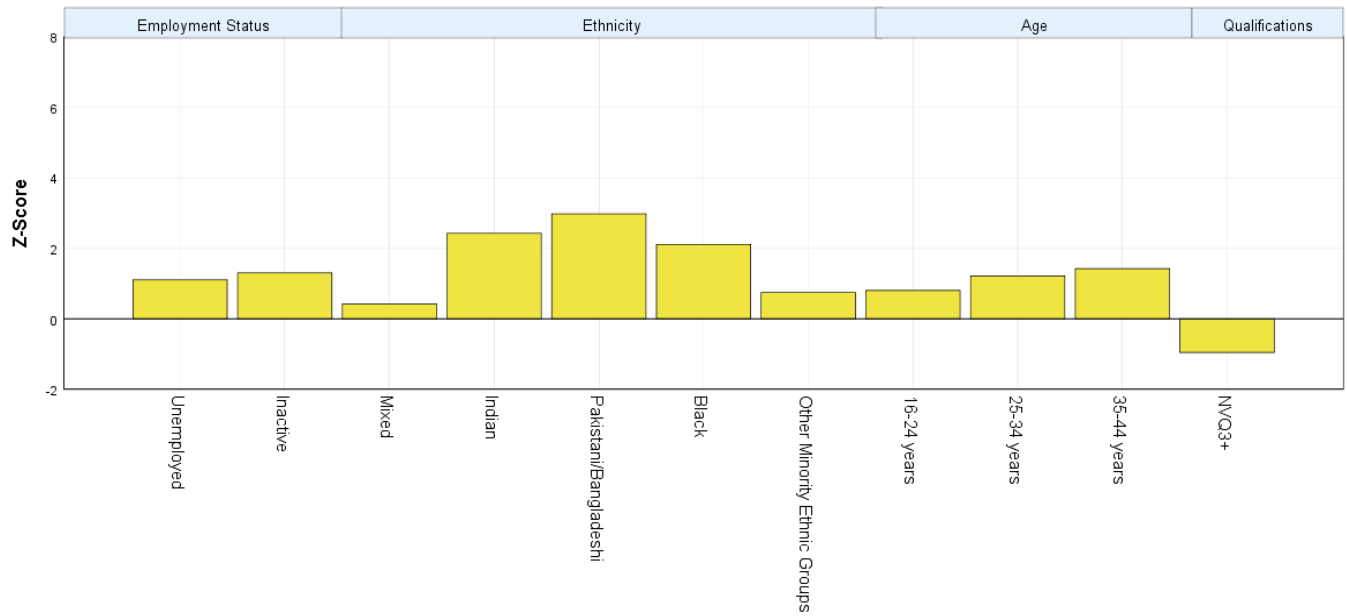

**Figure 9** Metropolitan Minority Struggle Profile

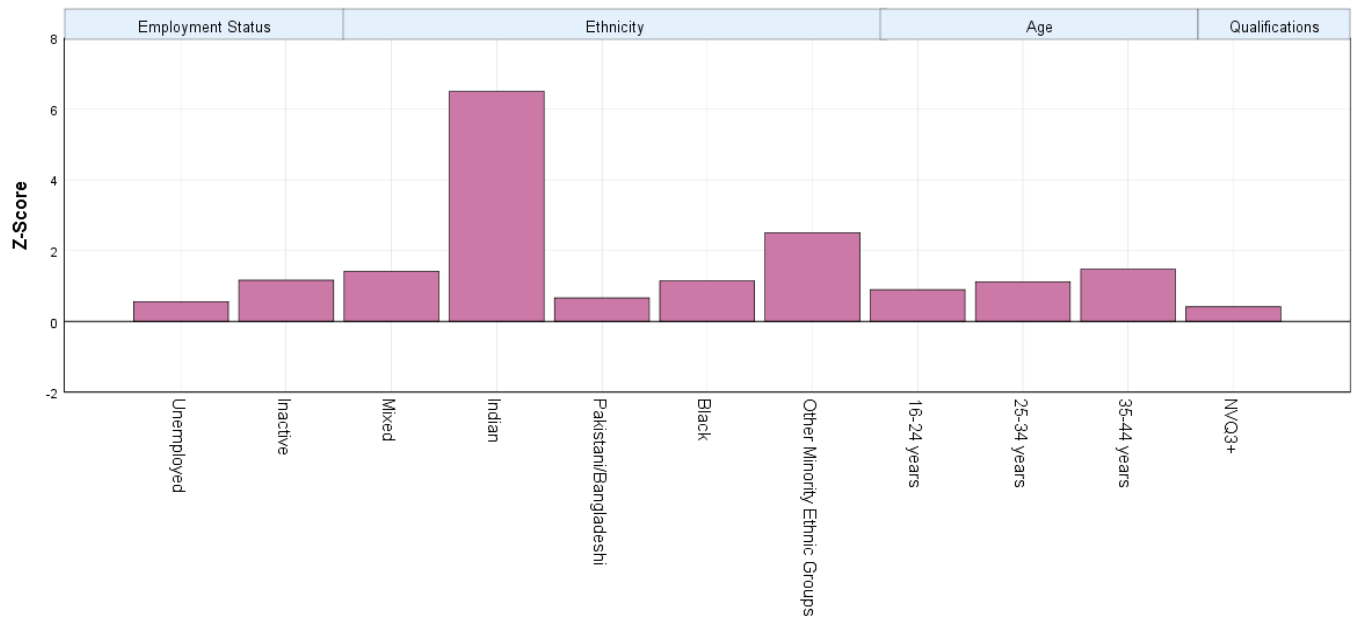

**Figure 10** Indian Metropolitan Living Profile

## Cluster 5 - Ethnically Diverse Career Climbers

This group (Figure 11) has an above average mix of Black, Mixed and other minority ethnic groups, except Pakistani-Bangladeshi and Indian ethnicity which are less prevalent in this and the Student Central cluster. A large proportion of residents are aged 35 to 44 years and are well-educated, with most awarded NVQ3+ qualifications. Ethnically Diverse Career Climbers areas include Milton Keynes and much of North and East central London.

## Cluster 6 - Student Central

The Student Central group (Figure 12) contains above average unemployed and inactive residents with the highest above average proportion of 16 to 24-year olds across all clusters. Those aged 25 to 34 years are also above average. A mix of all majority and minority ethnicities exists with none significantly more dominant. Student Central areas include Leeds, Lancaster, Warwick and Edinburgh.

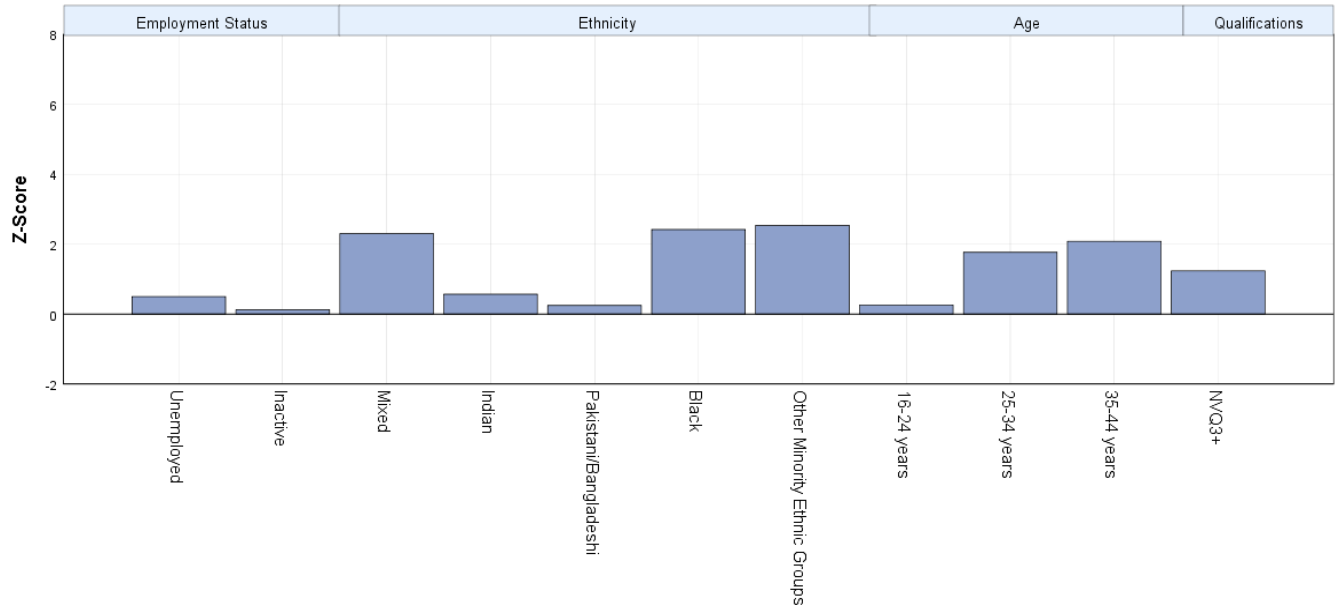

**Figure 11** Ethnically Diverse Career Climbers Profile

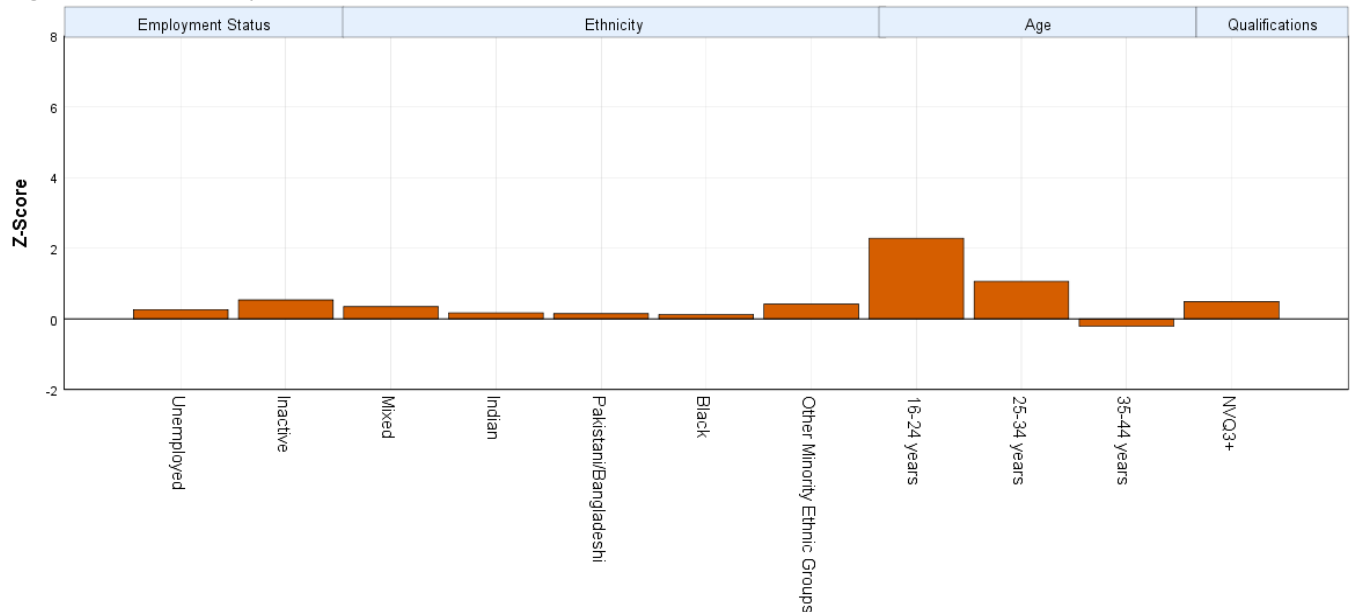

**Figure 12** Student Central Profile

## Cluster 7 - Pakistani-Bangladeshi Inequality

Another at-risk digital accessibility group (Figure 13) is the Pakistani-Bangladeshi Inequality cluster. Here areas have above average residents aged 25 to 34 years old and the highest proportion of residents of Pakistani-Bangladeshi ethnicity. Most do not have qualifications surpassing NVQ3+ and are inactive and unemployed. Internet access could potentially aid employment prospects. Pakistani-Bangladeshi Inequality areas generally located in Northern England, including Bradford, Oldham and Luton, with Tower Hamlets the only Southern district.

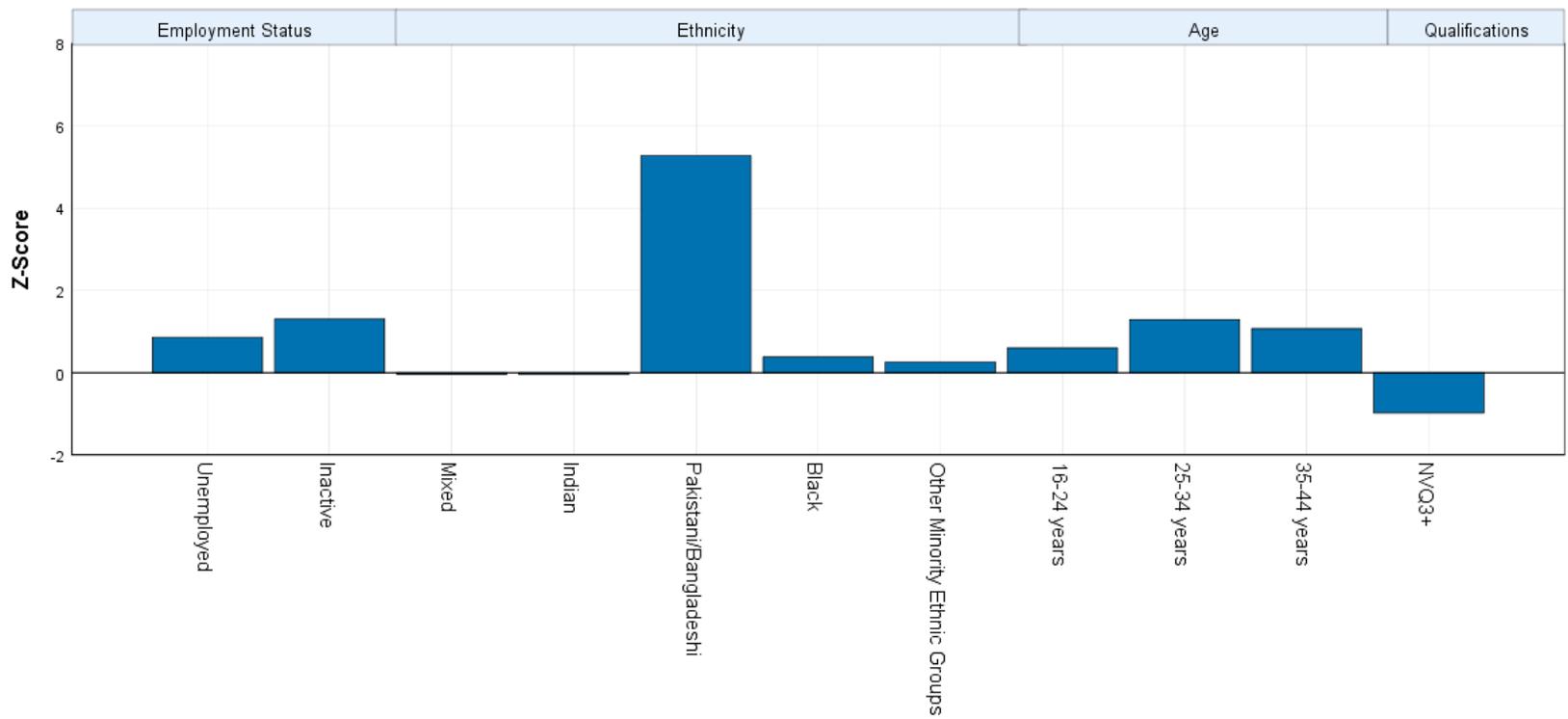

**Figure 13** Pakistani-Bangladeshi Profile

# Cluster Effectiveness

When analysing cluster outputs, the ANOVA F number, number of cases in each cluster and boxplots are known determiners of cluster effectiveness.

Table 2: ANOVA F Results

| Z-Scored Variables | F |
|---|---|
| Unemployed | 28.5 |
| Inactive | 42.8 |
| Mixed | 52.0 |
| Indian | 125.1 |
| Pakistani/Bangladeshi | 162.9 |
| Black | 90.8 |
| Other Minority Ethnic Groups | 103.5 |
| Aged 16-24 years | 136.2 |
| Aged 25-34 years | 67.6 |
| Aged 35-44 years | 51.8 |
| NVQ3+ | 45.0 |

The ANOVA F value denotes which variables contribute most to the resultant clusters (Sarstedt & Mooi, 2019). In Table 2, all F values are high, have similar values, averaging 82.4, and have a maximum variation of 134.4, which shows all variables contribute to cluster categorisation and no single variable dominates, nor do any variables offer little or no discrimination. The highest contributing variables are Pakistani/Bangladeshi, Indian, other minority ethnic groups and those aged 16 to 24 years. The lowest contributing variables are unemployed, inactive, National Vocational Qualifications (NVQ) 3+, aged 35 to 44 years and Mixed ethnicity.

Table 3: Number of Cases/ Districts in each Cluster

| Cluster | Cases |
|---|---|
| Rural Retirees | 137 |
| Advancing White Professionals | 151 |
| Metropolitan Minority Struggle | 9 |
| Indian Metropolitan Living | 4 |
| Ethnically Diverse Career Climbers | 25 |
| Student Central | 38 |
| Pakistani-Bangladeshi Inequality | 6 |

Similarities in mean cluster centre distances (particularly clusters 4, 6 and 7) and extreme values exist (Figure 5). Table 3 highlights where differences exist between clusters in terms of sociodemographic characteristics and distribution of individual districts within clusters. Cluster 2 has the greatest number of districts, 147 more than the lowest number of districts in Cluster 4. The first two clusters contain the largest quantity of districts. Although an even spread of districts in clusters is optimum (Sarstedt & Mooi, 2019), districts across Great Britain are likely to have variation and some characteristics will be present in more districts than others due to the changing sociodemographic nature of this research. All 370 districts across England, Scotland and Wales are accounted for effectiveness (Everitt, et al., 2011).

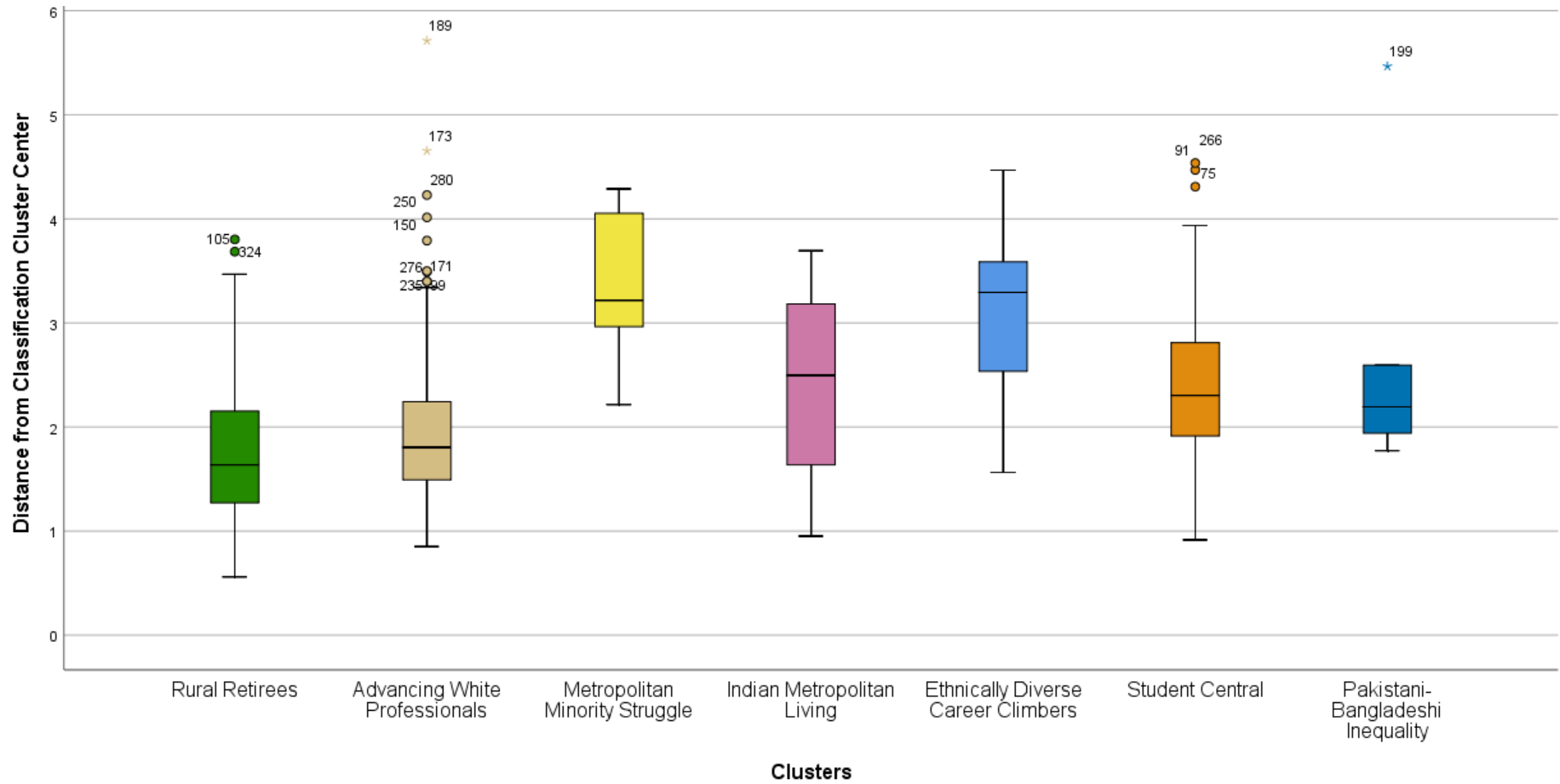

**Figure 5** Digital Accessibility Classification Boxplo

Figure 5 (boxplots) denotes overall cluster effectiveness. The mean cluster centre distances vary between 1.5 and 3.5. Most (3 out of 7 clusters) have mean cluster centre distances between 2.1 and 2.5. Cluster centres differ as every district in each cluster is likely to have differing levels of each of the 11 sociodemographic variables. Clusters form when districts have sufficiently similar variable levels to group (Kaufman & Rousseeuw, 1990).

Clusters 1, 2, 6 and 7 have outliers, 3, 4 and 5 do not. Cluster 2 has the most outliers and the highest overall outlier at 6.7. Clusters 1 and 6 contain the most variation, with the highest upper extreme and lowest lower extreme values, perhaps unsurprising since retirees and students are likely a diverse mix of people. Despite these variations, overall mean distances, extremes and outliers tend to be relatively low, a sign of better district-cluster suitability. Clusters 3, 4 and 7 represent poorly educated, inactive or unemployed residents of minority ethnicities, most at risk of digital inaccessibility.

Analysis of variable groupings and cluster effectiveness imply clusters at greatest risk of digital inaccessibility are 'Metropolitan Minority Struggle', 'Indian Metropolitan Living' and 'Pakistani-Bangladeshi Inequality'. Those in clusters at least risk of digital inaccessibility are 'Rural Retirees', 'Advancing White Professionals', 'Ethnically Diverse Career Climbers' and 'Student Central'.

## External Validation

To evaluate the digital accessibility classification and validate pen portraits, Great Britain-wide Ofcom (2019a) Telecommunications Operator Performance data, followed by cluster case study-specific analysis with ONS (2019b) Internet Users data was undertaken. Ofcom data refers to wireless mobile internet access, a good access indicator where wired is unavailable. Download and upload speed, used in calculating internet speed, was used at district level using all internet line types for validation. Download speed indicates speed data is obtainable from a server (e.g. video streaming), whereas upload speed determines how fast data is sent to others (e.g. sending emails) (Riddlesden & Singleton, 2014).

## 4.4.1 Ofcom Upload Performance

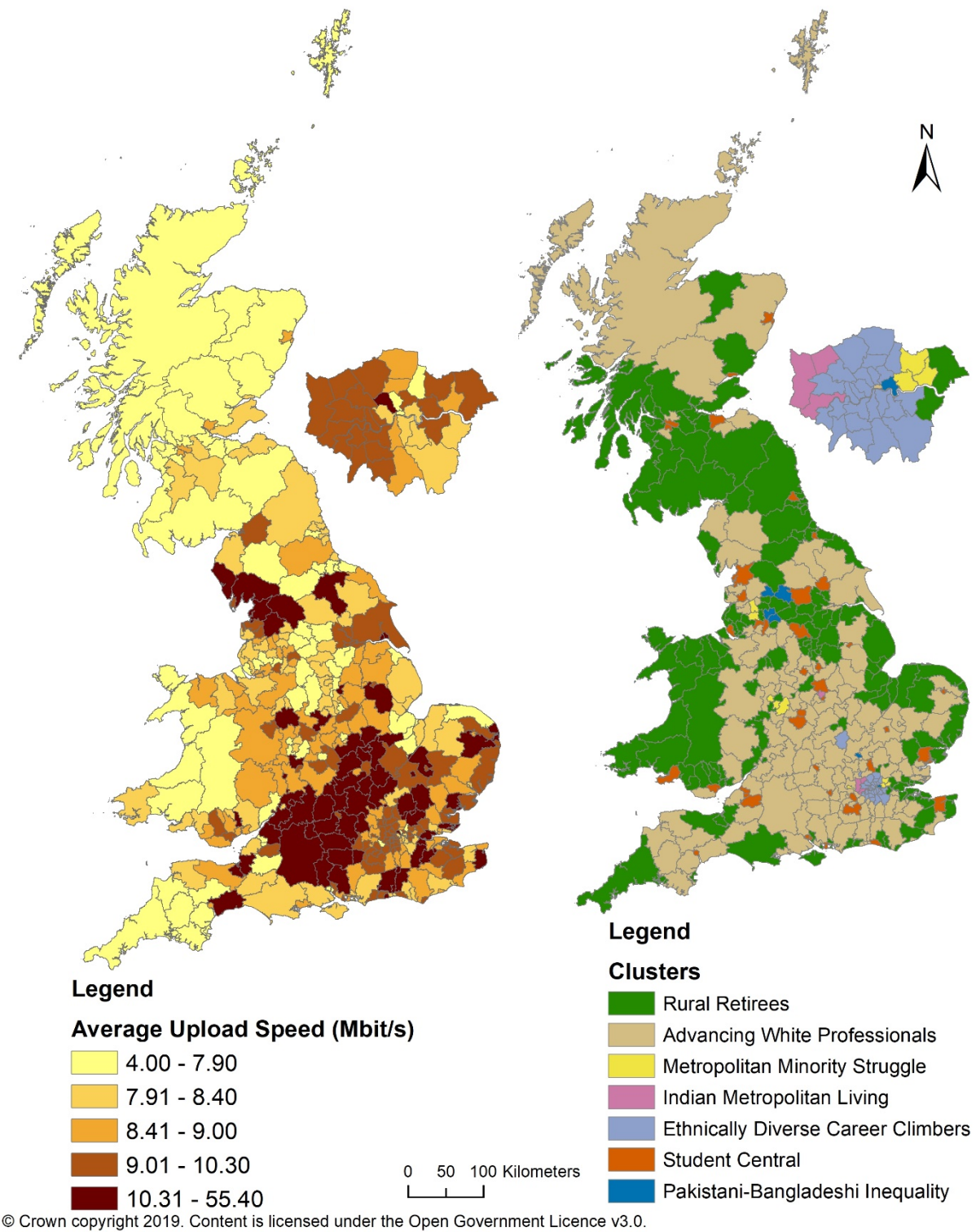

**Figure 14** Ofcom upload speed compared against the Digital Accessibility Classification

Figure 14 shows Ofcom (2019a) average upload speed alongside the Digital Accessibility Classification. Great Britain average upload speed is 'decent' at 10 Mbit/s (Ofcom, 2019c).

Highest upload speeds (> 10.31 Mbit/s) follow the 'Advancing White Professionals' cluster in locations including Wiltshire, West Oxfordshire and Mid-Suffolk. Lowest upload speeds (< 7.90 Mbit/s) follow the 'Rural Retirees' cluster in locations including mid-Wales, Cornwall and the Scottish Highlands. Clusters deemed to be at-risk from digital accessibility also show the lowest upload speeds and cover areas including Bradford ('Pakistani-Bangladeshi Inequality' cluster), Wolverhampton ('Metropolitan Minority Struggle') and Leicester ('Indian Metropolitan Living'). Those in the 'Ethnically Diverse Career Climbers' and 'Student Central' clusters show a range of high upload speeds, the majority > 9.01 Mbit/s, the second highest upload speed range. Outside of London in Milton Keynes the highest upload speeds are present, within Greater London upload speeds vary. London is geographically and demographically distinct (Dean, et al., 2012). Therefore, analysis and validation are focussed on the nationwide results rather than London-centric differences.

## 4.4.2 Ofcom Download Performance

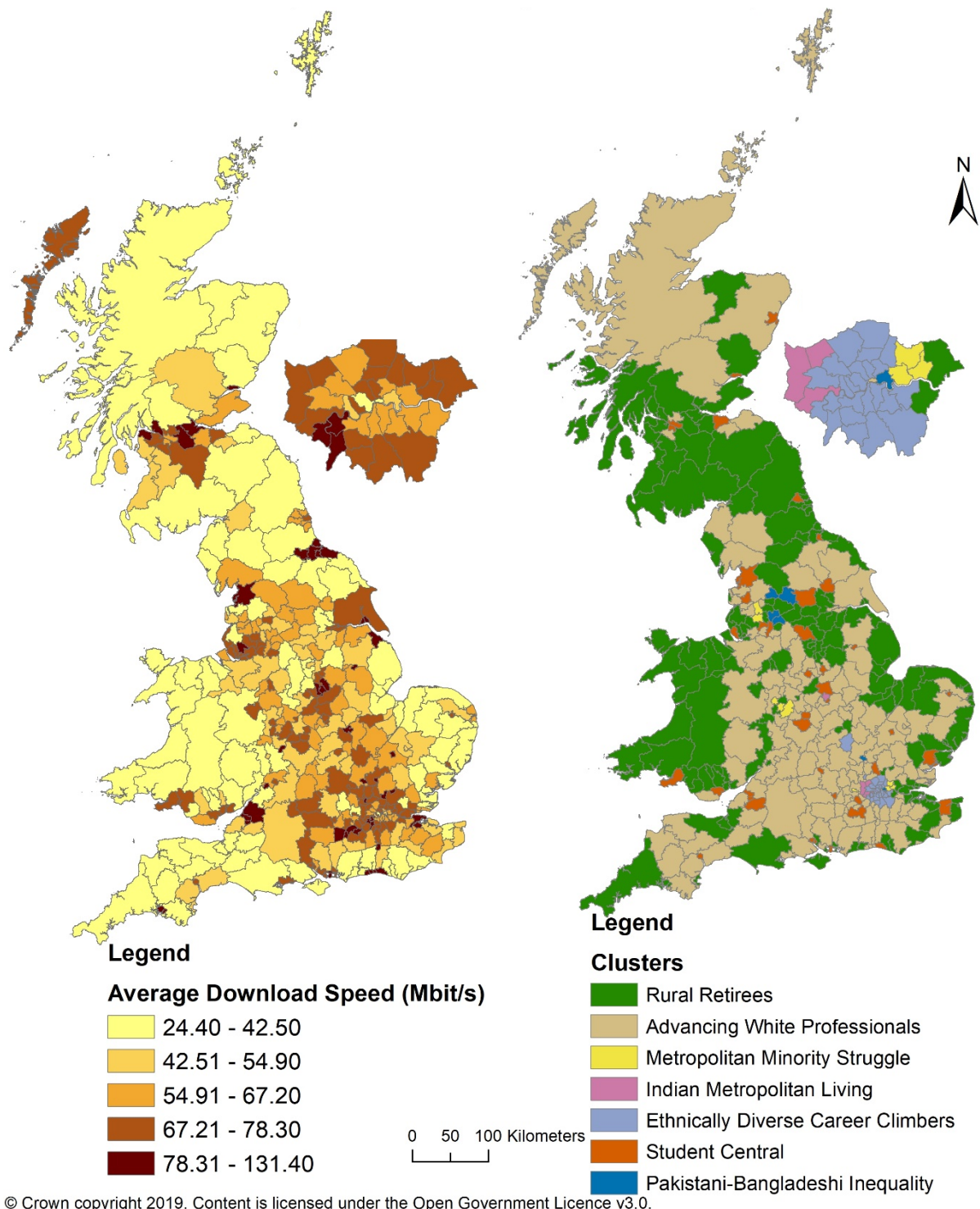

**Figure 15** Ofcom telecommunication operator download speed compared against the Digital Accessibility Classification

Figure 15 shows Ofcom (2019a) average download speed alongside the Digital Accessibility Classification. Great Britain's average download speed is 'superfast' at 58 Mbit/s (Ofcom, 2019c). Most of the country accesses the lowest average download speeds between 24.40 Mbit/s and 42.50 Mbit/s, thus determining clear cluster differences. Smaller spatial scale case study analysis would highlight these differences more clearly. In not at-risk areas similarities exist with 'Rural Retirees' experiencing the lowest upload and download speeds (< 42.50 Mbit/s). All other not at-risk clusters have high, above average download speeds. All in the Student Central cluster have high speeds above 54.91 Mbit/s. The 'Ethnically Diverse Career Climbers' cluster has higher speeds, averaging 67.40 Mbit/s across areas. Speeds for the 'Advancing White Professionals' vary but the majority surpass 54.91 Mbit/s.

At-risk area download speeds vary, some show above the Great Britain average and some below average. 'Indian Metropolitan Living' and 'Metropolitan Minority Struggle' have high download speeds between 67.21 and 78.30 Mbit/s. Pakistani-Bangladeshi areas vary either side of the Great Britain average between 54.91 and 67.20 Mbit/s. One third of areas are below the Great Britain average.

Ofcom data validates the classification nationwide. Generally, assumptions are validated: rural clusters and socio-demographically disadvantaged clusters have less digital access. However, some variation exists, particularly clusters in cities. Those disadvantaged in at-risk clusters, as in Xiang et al (2018) Central Beijing educational inequality work, should be validated further using case studies (Table 4) enhancing clarification and pinpointing areas within largely urban clusters most requiring additional digital access help.

## Case Studies

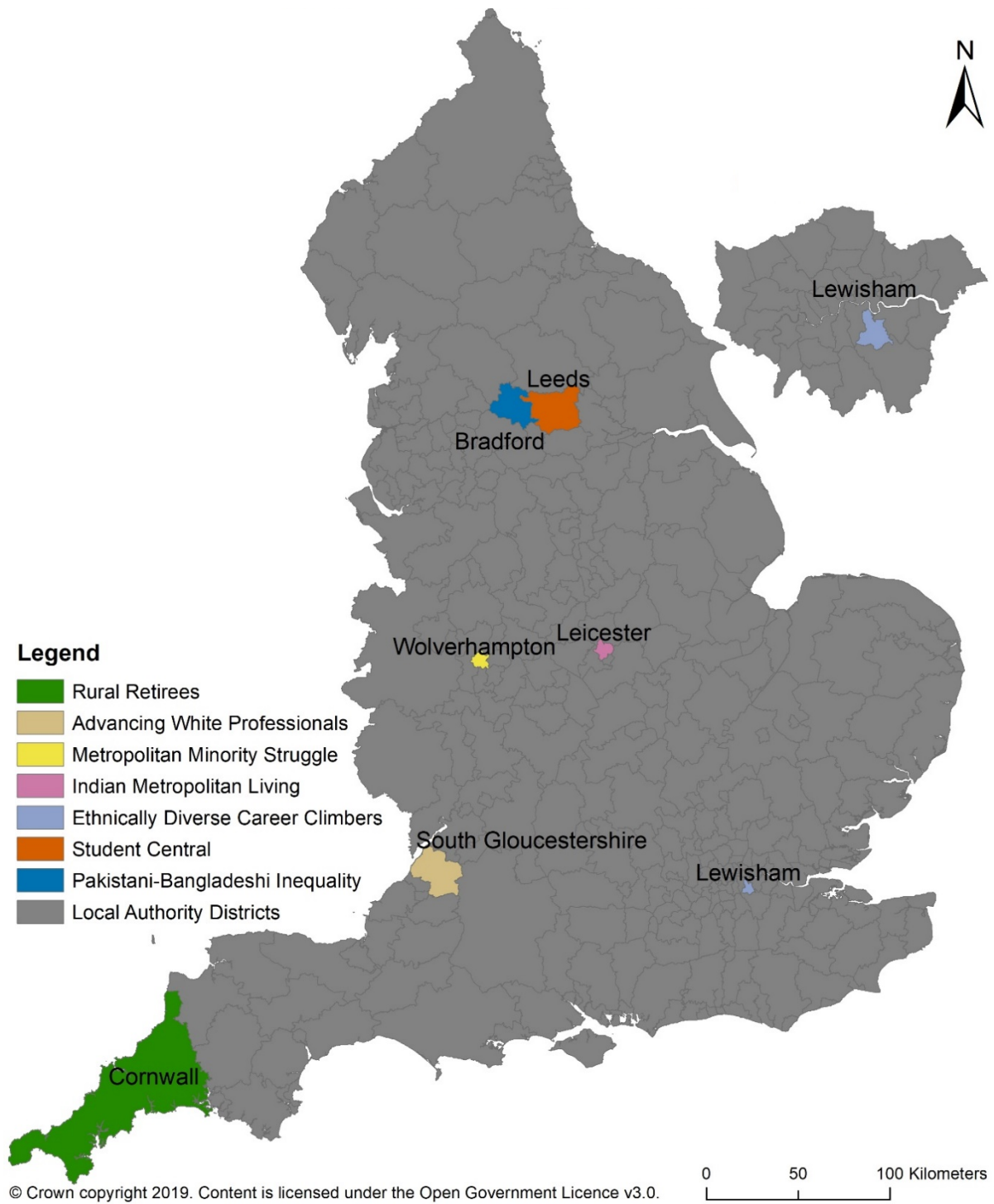

**Figure 16** Case study locations

Figure 16 presents one stereotypical district from each Classification cluster as seen in the ONS APS Dataset (2019a).

Table 4: Rank of UK Internet Users using Example At-Risk and Not At-Risk areas from Digital Inaccessibility Classification. Derived from ONS (2019b)

| Digital Accessibility Risk | | Used in Last 3 months | Never Used/ 3+ months since use |
|---|---|---|---|
| Not At Risk | Cornwall | 25th | 29th |
| Not At Risk | South Gloucestershire | 12th | 42nd |
| Not At Risk | Lewisham | 15th | 99th |
| Not At Risk | Leeds | 6th | 64th |
| At Risk | Bradford | 45th | 36th |
| At Risk | Leicester | 73rd | 66th |
| At Risk | Wolverhampton | 138th | 23rd |
| There are 174 total areas in the NUTS scale. | | | |

Table 4 presents case study districts ranked by recent internet usage. ONS internet usage data is to a NUTS level scale, a coarser spatial scale and not equivalent to districts (Longley, 2012). Generally, this would not be usable as validation. However, the cluster case studies have the same physical geographic boundaries thus results are relatively comparative.

Table 4 highlights the difference in internet usage by at-risk and not at-risk areas. The not at-risk 'Ethnically Diverse Career Climbers' cluster in Lewisham has the lowest 'never used' or 'used over 3 months' nationwide internet usage at 99th, most residents are connected and regular internet users. The highest internet usage over the last 3 months is 'Student Central', Leeds, at 6th. Other not at-risk clusters also have high internet usage: South Gloucestershire, an 'Advancing White Professionals' cluster, has the next highest internet usage, closely followed by 'Ethnically Diverse Career Climbers' in Lewisham and Cornwall, a 'Rural Retirees' cluster. All are within the top 25 internet users.

In contrast, at-risk areas are ranked lower in internet usage over the last 3 months. Wolverhampton, a 'Metropolitan Minority Struggle' cluster, is ranked lowest for internet usage in the last 3 months against the other case study areas and ranked 138th out of 174 UK areas. Wolverhampton is ranked highest regionally and nationwide in the at-risk clusters for those who have never accessed the internet or not accessed in over 3 months. Leicester, an 'Indian Metropolitan Living' cluster, has next lowest 3-month internet usage followed by Bradford. Clear differences exist between case study clusters. At-risk clusters have more residents not having used the internet and the lowest level of internet users in the last 3 months.

'Metropolitan Minority Struggle', 'Indian Metropolitan Living' and 'Pakistani-Bangladeshi Inequality' clusters are at greatest risk of digital inaccessibility in the Digital Accessibility Classification. This research is further verified by Ofcom upload and download speeds and in specific UK Internet Usage case studies. The Great Britain upload speeds mapped alongside the classification supports several classification findings. A physical digital divide exists with 'Rural Retirees' receiving the slowest upload speeds and urban 'Advancing White Professionals' receiving the highest upload speeds. Sociodemographic differences exist with all disadvantaged clusters associated with digital inaccessibility, obtaining the lowest upload speeds. By contrast, all privileged clusters access the highest upload speeds. Findings are validated across Great Britain. However, variation does exist in London where upload speeds vary and districts are not consistent with the digital accessible-inaccessible cluster divisions.

Validation of download speeds is less definitive with the majority of speeds across Great Britain relatively low. Clear, immediate differences between classifications deem digital accessible and inaccessible clusters difficult to determine. The physical access division exists to an extent with Metropolitan clusters ('Indian Metropolitan Living' and 'Metropolitan Minority Struggle') generally having high download speeds. Sociodemographic cluster divisions are less clear, with variations in download speeds. For instance, one-third of the digitally inaccessible Pakistani-Bangladeshi cluster has low download speeds, below the Great Britain average. The lack of clear-cut validations among all clusters in Figure 15 are in part due to the wide spatial scale the Great Britain download speeds cover. The additional case study specific validation further clarifies classification findings.

Ranked UK internet users (Table 4) further corroborate classification findings. Digitally inaccessible clusters correspond with areas in Table 4 with fewer regular internet users, greater periods between internet use and more residents who have never used the internet before. The reverse applied for people in clusters deemed to have greater digital accessibility.

# Conclusion

To conclude, this innovative, solely sociodemographic geodemographic Digital Accessibility Classification highlights districts across Great Britain at-risk and not at-risk from digital inaccessibility. Lack of effective digital access in disadvantaged districts can impact employment, health, education and more. This classification surpasses previous digital access studies as data specifically highlights those disadvantaged. The data is the most recent release, open source and covers Great Britain and the methodology is transparent, allowing scrutiny.

Nationwide and case study validations have supported the literature-guided Classification findings. Upload speed validation supports physical digital access divisions impacting rural areas with low internet speeds, particularly in the Scottish Highlands, Mid-Wales, the Scottish borders and in the South West of England. Distinct differences exist between at-risk and not at-risk clusters. Case-study specific classification validation against internet usage supports that residents who experience low upload speeds, low internet usage and who have not used the internet in over 3 months (if connected to the internet at all), align strongly with internet inaccessibility clusters from the Digital Accessibility classification. These residents, found in the 'Metropolitan Minority Struggle', 'Indian Metropolitan Living' and 'Pakistani-Bangladeshi Inequality' clusters, exhibit above average sociodemographic variables associated with disadvantage. By contrast, in districts where residents receive high upload speeds, high internet usage and have a low proportion of people who have never used the internet, privileged sociodemographic variables are present. These areas not requiring targeted digital access help comprise the 'Advancing White Professionals', 'Rural Retirees', 'Student Central' and 'Ethnically Diverse Career Climbers' clusters.

# Classification Limitations
## Statistical Limitations

Statistics to determine the number of clusters, the Gap Statistic and Clustergram were run 500 and 100 times respectively. Number of iterations was determined from previous geodemographic classification literature (Xiang, et al., 2018; Singleton, et al., 2020). For the K-Means classification this could be increased, however, this is unlikely to change the number of clusters identified as most suitable nor the K-Means classification F number, number of cases per cluster or any other outcomes.

## Upload Validation Limitations

Upload speed validation supports overall classification findings. Greater London variations do not consistently match against cluster divisions; however, London is known to be demographically distinct (Dean, et al., 2012). Rerunning of the classification to analyse specific locations, such as London, may further distinguish differences.

Additional rerunning at a smaller spatial scale (e.g. postcode) could enable more specific digital access and resource targeting. Ofcom validation data is freely available at postcode level, however ONS data is not. Accessing secure ONS data to postcode level would lead to more specific spatial analysis, however without all data being freely, publicly available the classification cannot be scrutinised in-depth. Yet with postcode level ONS data released publicly, individuals most socially disadvantaged and vulnerable could potentially be identified and be at-risk from targeting by commercial and criminal organisations. This is the overarching reason for maintaining data to district level, allowing a high spatial resolution over Great Britain while being able to highlight specific districts at-risk from digital inaccessibility.

## Download Validation Limitations

Download speed validation less definitively supports classification findings. There is support for a physical digital access divide with higher download speeds in Metropolitan clusters compared to generally rural-based clusters. However, download speeds are low across the majority of Great Britain, specific differences between sociodemographic digital accessible and inaccessible clusters are difficult to determine. Specific case study validation strengthens findings.

## Internet User Validation Limitations

The case study validation used data from a NUTS level Internet User survey. Generally, this would not be comparable with the district level ONS data for the classification. NUTS has a coarser spatial scale than district. However, for these case studies, the same physical geographic boundary exists. Other case studies may not be comparable at the same geographic level. Care must be taken if this validation is required for other areas.

# Classification Reusability

## Short-Term Reusability

The Digital Accessibility research identifies at-risk clusters and their spatial extent to district level, which should aid future government policy. Districts requiring additional help could be targeted. Locations of inaccessibility are places where future infrastructure planning (e.g. 5G masts) or computer skills training would be beneficial. The sociodemographic variables selected are multidimensional, capturing key parts of society.

## Longer Term Reusability

The open source publicly available data and clear, transparent methodology used have enabled this in-depth analysis and break down of the classification. Methodology transparency allows the classification to be updated when new data releases are available and to sharper spatial scales using finer sociodemographic variable data, if local classifications are required to identify at-risk postcode zones for local councils. Data backlog and resource reallocation due to the Coronavirus pandemic have already seen ONS (2020) data releases delayed. The aforementioned delay, alongside what is expected to be the final nationwide census in 2021 means that this Digital Accessibility classification, a measure of social disadvantage and digital inaccessibility, is an important asset. It is perhaps one of the most up to date sociodemographic-focussed data classifications for a prolonged period, in a time when those who are already disadvantaged require the greatest support.

# Summary

The Digital Accessibility Classification successfully pinpoints districts across Great Britain requiring additional help to get connected to the internet or help gaining computer skills. Skill and knowledge improvement allow the internet to be used effectively to provide benefit and opportunities. In Great Britain, places including Bradford, Wolverhampton and Leicester are within the Pakistani-Bangladeshi Inequality, Indian Metropolitan Living and Metropolitan Minority Struggle clusters that require additional help. A total of 19 districts require immediate help overcoming the sociodemographic impacts of digital inaccessibility. Specific sociodemographic variables tend to be associated with social disadvantage and digital inaccessibility. It is these variables which local and national government need to focus on and tailor solutions towards. The entire population, regardless of age, ethnicity, employment status and qualifications should be able to access the internet and be taught the skills to use it.